\documentclass{ws-procs9x6}

\begin{document}

\title{TWISTED MODULI AND SUPERSYMMETRY BREAKING}

\author{S.~F.~KING}

\address{Department of Physics and Astronomy, \\
University of Southampton, Southampton, SO17 1BJ, U.K. \\
E-mail: sfk@hep.phys.soton.ac.uk}

\author{D.~A.~J.~RAYNER}

\address{Dipartimento di Fisica `G. Galilei',
Universit\'{a} di Padova and INFN, \\
Sezione di Padova, Via Marzola 8, I-35131 Padua, Italy\\
E-mail: rayner@pd.infn.it}

\maketitle

\abstracts{We discuss how localized twisted moduli in type I string
theory can provide a string realization of brane world supersymmetry
breaking models.}

\begin{figure}[th]
\centerline{\epsfxsize=4.0in\epsfbox{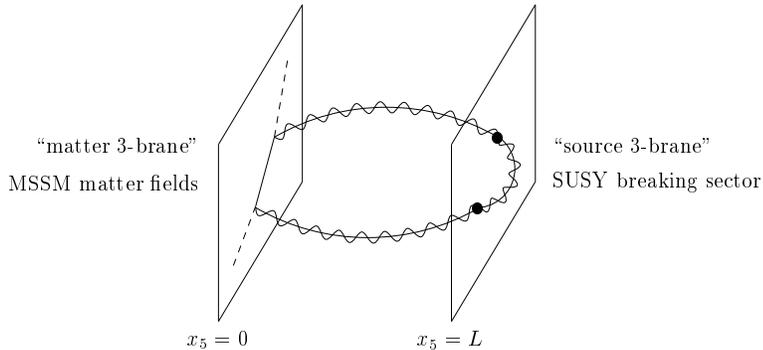}}   
\caption{A generic parallel-brane setup where MSSM fields are
sequestered away from SUSY-breaking along an extra dimension $x_{5}$.}
 \label{fig:generic}
\end{figure}

String theory offers the most appealing
framework for unifying all four fundamental forces together.  Much
recent work has been devoted to studying the problem of supersymmetry
(SUSY) breaking in the context of higher-dimensional models involving
parallel branes~\footnote{Here we use the term ``brane'' to describe
any subspace of the higher-dimensional theory.}.  These models are
inspired by the Horava-Witten 
setup~\cite{hw}, but studied using effective field theory (EFT)
techniques with an ultraviolet cutoff that is often identified with
the string scale $M_{\ast}$.  Generic models
confine the MSSM fields to a 3-brane which is spatially-separated
(sequestered) along a
transverse extra dimension from another 3-brane where 
SUSY is broken by a vacuum expectation value (VEV) $F$.  Direct
couplings between sectors are assumed to arise only from the exchange
of bulk fields with masses {\it above} the string scale $M_{\ast}$.
After integrating out these heavy modes, we find that direct couplings
(i.e. scalar masses $m_{\phi}$) are exponentially suppressed by the
separation, R
\begin{eqnarray}
 m_{\phi , tree}^{2} \sim e^{- M_{\ast} R} \, \frac{F^{2}}{M_{\ast}^{2}}
\end{eqnarray}
Hence, SUSY breaking is (predominantly) mediated across the bulk via
the superconformal anomaly~\cite{rs} or gaugino loops~\cite{gaugino}, so that
the leading contributions to scalar masses arise at 1-loop as shown in
Figure \ref{fig:generic}. 

These types of models are phenomenologically appealing since they
alleviate the SUSY flavour problem by suppressing 
flavour-changing neutral-currents (since loop corrections are
flavour-blind), and also provide a simple and
elegant realization of hidden-sector scenarios.  In traditional
4D models, SUSY breaking originated in a hidden-sector that was only {\it
hidden} from the visible MSSM fields by the weakness of direct
couplings, e.g. Planck-scale suppressed couplings in gravity mediated
models.  In contrast, extra-dimensional theories provide a geometric way
to separate the two sectors such that they occupy completely
distinct and (apparently) unrelated physical spaces!  

However, there has been recent criticism of these {\it sequestered}
models that question whether the required form of the K\"{a}hler
potential (essentially of the no-scale type) can be realized in generic string
compactifications~\cite{dine}.  These authors dispute the original
claim of Ref.~\cite{rs} that direct couplings between sectors are suppressed,
by arguing that the exchange of bulk modes with masses {\it below} the
string scale $M_{\ast}$ do not miraculously cancel, but instead induce
inter-brane couplings that generate non-trivial scalar masses at
tree-level.  Nevertheless, the sequestering mechanism is still highly
attractive which has motivated our attempts to embed sequestering into
type I models~\cite{us} using twisted moduli~\cite{benakli}.

Twisted moduli ($Y_{k}$) are closed string states that become trapped at
orbifold fixed points during compactification.  They play an important
role in cancelling anomalies in type I theory through a generalized
Green-Schwarz mechanism~\cite{gs}.  They appear in the gauge
kinetic function at tree-level (e.g. $f_{9} = S + \sum_{k} Y_{k}$ for
a gauge group arising from a D9-brane) and therefore provide a new
source of gaugino masses if there auxiliary F-terms acquire non-zero
VEVs~\cite{benakli}.  More recently, they have been shown to be
crucial for dilaton stabilization in type I models using only a
{\it single} gaugino condensate~\cite{abel}.  Their localization at 4D
orbifold fixed points implies that they only couple to string states
that overlap with them, and so if MSSM states
are localized at a distance from the twisted moduli, we anticipate a
suppression of direct couplings.  

\begin{figure}[th]
\centerline{\epsfxsize=4.0in\epsfbox{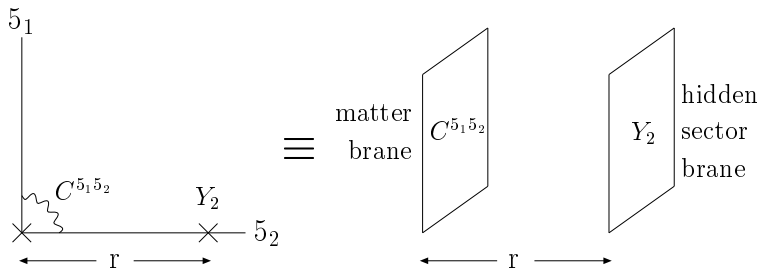}}   
\caption{We propose an equivalence between intersecting D-brane
constructions and parallel brane EFTs, where the matter (hidden
sector) brane is like open string states $C^{5_{1} 5_{2}}$ (twisted
moduli $Y_{2}$) respectively.}
 \label{fig:equiv}
\end{figure}

In Figure ~\ref{fig:equiv}, we propose that parallel brane EFTs are
equivalent to string constructions involving intersecting D5-branes and
twisted moduli.  Matter fields are
identified with open string states that stretch between the D-branes
and are effectively localized at the 4D intersection point, while the
hidden sector brane is like a twisted moduli localized at a fixed
point away from the origin.  We can also include the contributions from
dilaton $S$ and untwisted moduli $T_{i}$ fields that live in the full
10D space and 
provide a string theoretic description of gravity mediation.  In the
limit that SUSY breaking originates from the twisted moduli F-term VEV,
we expect that scalar masses ($m_{C^{5_{1} 5_{2}}}$) have
an explicit dependence on the distance $r$
from the SUSY breaking.  However, using the standard
heterotic-inspired K\"{a}hler potential for open string states
($C^{5_{1} 5_{2}}$) at the intersection between branes~\cite{ibanez98}, 
\begin{eqnarray}
 K_{C^{5_{1} 5_{2}}} = \frac{1}{\sqrt{S+\bar{S}} \,
  \sqrt{T_{3}+\bar{T_{3}}}} \, \left| C^{5_{1} 5_{2}} \right|^{2}
\end{eqnarray}
we discover that the
masses are completely independent of this separation.  This leads us to
postulate a modified K\"{a}hler potential for these sequestered states
\begin{eqnarray}
 K_{C^{5_{1} 5_{2}}}' = exp \left[ \frac{1}{6} (1-\eta) \, (Y_{2}+
  \bar{Y_{2}})^{2} \right] \frac{1}{\sqrt{S+\bar{S}} \,
  \sqrt{T_{3}+\bar{T_{3}}}} \, \left| C^{5_{1} 5_{2}} \right|^{2}
\end{eqnarray}
where $\eta= e^{- M_{\ast} \, r}$ or $e^{- ( M_{\ast} \, r)^{2}}$.
Notice that we can recover the standard result in Eq. 2 by taking the
limit that $r \longrightarrow 0$.  
We attribute the suppression to either (i) the propagation of heavy
modes with masses above the string scale~\cite{rs,gaugino} ($\eta=
e^{- M_{\ast} \, r}$); or (ii) arising from non-perturbative
worldsheet instanton corrections~\cite{altexp} that are interpreted
as (massive) strings stretched between the different fixed points
($\eta= e^{- (M_{\ast} \, r)^{2}}$). 

Using our modified K\"{a}hler potential, we use a model-independent
parametrization of F-terms involving Goldstino
angles~\cite{ibanez98,ibanez94} to derive general expressions for
trilinears and scalar masses, where different limits of these
Goldstino angles correspond to dilaton ($S$), moduli ($T_{i}$) or
twisted moduli ($Y_{k}$) dominated SUSY breaking.  We anticipate that
our general expressions can be used to extend previous type I theory
parameter scans~\cite{allanach}, but incorporating both gravity and
gaugino mediated contributions.


\end{document}